\documentclass[12pt]{article}
\usepackage{epsfig}
\usepackage{pst-plot,url,epsf}
\usepackage{axodraw}
\newcommand{\beq}{\begin{equation}}
\newcommand{\eeq}{\end{equation}}
\newcommand{\bea}{\begin{eqnarray}}
\newcommand{\eea}{\end{eqnarray}}

\newcommand{\epm}{e^+e^-}

\newcommand{\ra}{\rightarrow}
\def\earr{\end{array}}
\def\barr#1{\begin{array}{#1}}

\begin{document}
\thispagestyle{empty}
\begin{flushright}
December 2003\\
\vspace*{1.5cm}
\end{flushright}
\begin{center}
{\LARGE\bf An anomalous $Wtb$ coupling at a linear collider\footnote{Work 
           supported in part by the Polish State Committee for Scientific 
           Research (KBN) under contract No. 2~P03B~045~23.}}\\
\vspace*{2cm}
K. Ko\l odziej\footnote{E-mail: kolodzie@us.edu.pl}\\[1cm]
{\small\it
Institute of Physics, University of Silesia\\ 
ul. Uniwersytecka 4, PL-40007 Katowice, Poland}\\
\vspace*{4.5cm}
{\bf Abstract}\\
\end{center}
Differential cross sections of secondary particles in a 
process of top quark pair production and decay into six fermions 
at a linear collider with an unpolarized and a longitudinally polarized 
electron beam are computed to the lowest order in the standard model
and in the presence of an anomalous $Wtb$ coupling. It
is illustrated that the latter has a little impact on
the differential cross sections.
In particular, it is shown that the angular distribution of a secondary 
lepton receives practically no contribution from the anomalous $Wtb$ coupling, 
even if the
top quark is produced off shell and the non-double resonance
background contributions are taken into account. This finding is
in accordance with 
the decoupling theorem that has been proven in literature \cite{GHR}
in the narrow top quark width approximation.

\vfill

\newpage

\section{Introduction.}

Future high luminosity $\epm$ linear collider with its very
clean experimental environment will be the most suitable tool
for searching for the effects of physics beyond the standard model (SM).
As the top quark is the heaviest matter particle ever observed, 
with mass close to the energy scale of the electroweak symmetry breaking,
such non-standard effects
may in particular manifest themselves in departures
of the top quark properties and interactions from those predicted by SM.
Needless to say, the discovery of such departures would give hints 
toward understanding physics beyond SM at higher energy scales.
Therefore, measurements of the top quark properties and interactions,
at the precision level of a few per mille, belong to the research 
program of any future $\epm$ linear collider \cite{NLC}.

At the linear collider, top quarks are produced in pairs 
in the process
\bea
\label{eett}
         e^+e^- \rightarrow  t \bar{t}.
\eea
Due to a large decay width, $t$ and $\bar{t}$ of reaction (\ref{eett}) 
almost immediately decay, predominantly through the following channels
\bea
\label{tbw}
t \ra bW^+ \qquad {\rm and} \qquad \bar{t} \ra \bar{b}W^-,
\eea
with $W^+$ and $W^-$ decaying into two fermions each.
Thus, what one actually observes are reactions of the form
\bea
\label{ee6f}
\epm \ra 6\,{\rm f},
\eea
where 6f denotes a 6 fermion  final state that is possible in SM.
Any specific channel of (\ref{ee6f}) receives contributions typically 
from several hundred Feynman diagrams already at the lowest order of SM, 
whereas there are only two signal diagrams that contribute to it,
for example, see Fig.~1a.

Another interesting consequence of the very short life time 
of the top, $\tau_t = 1/\Gamma_t \approx 1/(1.5 \;{\rm GeV})$, is that its 
spin can be directly observed. As the top quark decays before it hadronizes,
information about its spin is passed directly to its decay products and can be
best gained by measurement of the angular
distribution of its decay products \cite{Jezabek}.
Therefore, it is interesting to look at an influence, which extensions of the 
pure left-handed $Wtb$ coupling that governs decays (\ref{tbw}) to the
lowest order of SM, may have on some differential cross sections
of (\ref{ee6f}). This issue has been already investigated in literature 
\cite{effl}, \cite{Wtb}, \cite{CK1}, \cite{GHR}. In \cite{GHR}, 
a decoupling theorem has been proven,
which states that the angular distribution of the secondary lepton resulting
from a decay of top quark produced in (\ref{eett}) receives no contribution
from the anomalous $Wtb$ coupling
in the narrow top quark width approximation.

In the present article, the influence of the anomalous $Wtb$ coupling 
on the process of top quark pair production and decay is analyzed
numerically in a more realistic case, where the top quark pair is produced 
off shell and a complete set of the Feynman diagrams contributing to 
any specific channel of (\ref{ee6f}) at tree level
is taken into account, including the non-double 
resonance background contributions. It is illustrated
that the existence of anomalous form factors $f_2^{\pm}$ which assume
values within present experimental limits has rather little an
influence on differential cross sections of secondary particles of reaction 
(\ref{ee6f}). 
In particular, it is shown that the angular distribution of a secondary 
lepton of a semileptonic channel of (\ref{ee6f}) receives practically 
no contribution from the anomalous $Wtb$ coupling, 
which is in accordance with the decoupling theorem quoted above.

\section{An anomalous $Wtb$ coupling}

Departures of the $Wtb$ coupling from SM can be best parametrized in terms 
of the effective Lagrangian \cite{effl}. The effective Lagrangian used
in this paper has been written down in Eq.~(3) of \cite{CK}. The 
corresponding Feynman rules for the $Wtb$ vertex are
\bea
\label{Wtb1}
\barr{l}
\begin{picture}(90,80)(-50,-36)
\Text(-45,5)[lb]{$t, p_{t}$}
\Text(35,27)[rb]{$b, p_b$}
\Text(35,-27)[rt]{$W_{\mu}^+$}
\Vertex(0,0){2}
\ArrowLine(0,0)(35,25)
\ArrowLine(-45,0)(0,0)
\Photon(0,0)(35,-25){2}{3.5}
\end{picture} 
\earr
\barr{l}
\\[6mm]
\longrightarrow \quad
\Gamma^{\mu}_{t\ra bW^+}=-{g\over\sqrt{2}}V_{tb}\;
\left[\,\gamma^{\mu}\left(f_1^- P_- +f_1^+ P_+\right) \right.
                                                                   \\[3mm]
\quad \qquad \qquad \qquad \qquad \qquad \left.-i\sigma^{\mu\nu}
   \left(p_t - p_b\right)_{\nu}\left(f_2^- P_- +f_2^+ P_+\right)/ m_W\,\right],
\earr
\eea
and
\bea
\label{Wtb2}
\barr{l}
\begin{picture}(90,80)(-50,-36)
\Text(-45,5)[lb]{$\bar{t}, p_{\bar{t}}$}
\Text(35,27)[rb]{$\bar{b}, p_{\bar{b}}$}
\Text(35,-27)[rt]{$W_{\mu}^-$}
\Vertex(0,0){2}
\ArrowLine(35,25)(0,0)
\ArrowLine(0,0)(-45,0)
\Photon(0,0)(35,-25){2}{3.5}
\end{picture} 
\earr
\barr{l}
\\[6mm]
\longrightarrow \quad
\Gamma^{\mu}_{\bar{t}\ra \bar{b}W^-}=-{g\over\sqrt{2}}V_{tb}^*\;
\left[\,\gamma^{\mu}\left(\bar{f}_1^- P_- + \bar{f}_1^+ P_+\right) \right.
                                                                   \\[3mm]
\quad \qquad \qquad \qquad \qquad \qquad \left.-i\sigma^{\mu\nu}
   \left(p_{\bar{t}}-p_{\bar{b}}\right)_{\nu}\left(\bar{f}_2^- P_- 
                     +\bar{f}_2^+ P_+\right)/ m_W\,\right].
\earr
\eea
In Eqs.~(\ref{Wtb1}) and (\ref{Wtb2}), $V_{tb}$ is the element of
the Cabibbo-Kobayashi-Maskawa (CKM) matrix,
$P_{\pm}=(1\pm\gamma_5)/2$ are chirality projectors, $p_t$ ($p_{\bar{t}}$)
is the four momentum of the incoming $t$ ($\bar{t}$) and
$p_b$ ($p_{\bar{b}}$) is the four momentum of the outgoing $b$ ($\bar{b}$);
$f_{i}^{\pm}$ and $\bar{f}_{i}^{\pm}, \; i=1,2$ are the vertex form factors.
If the $Wtb$ interaction conserves $\cal{CP}$, which is
assumed in the rest of the article, then the following 
relationships between
the form factors of Eqs.~(\ref{Wtb2}) and (\ref{Wtb1}) hold
\beq
\label{rel}
\bar{f}_1^+=f_1^+, \quad \bar{f}_1^-=f_1^-, \qquad {\rm and} \qquad
\bar{f}_2^+=f_2^-, \quad \bar{f}_2^-=f_2^+.
\eeq
The lowest order SM vertex is then reproduced by setting 
\beq 
\label{sm}
f_{1}^{-}=1, \qquad f_{1}^{+}=f_{2}^{-}=f_{2}^{+}=0.
\eeq

As the experimental value of $\left|V_{tb}\right|$ is $0.9990$--$0.9993$ 
\cite{PDG} and deviation 
of the (V+A) coupling $f_{1}^{+}$ from zero is severely constrained by 
the CLEO data on $b\rightarrow s\gamma$ \cite{cleo}, in the 
numerical analysis, the results of which are presented in the next section, 
$f_1^{\pm}$ are fixed at their SM values
of Eq.~(\ref{sm}), $V_{tb}=1$ and modifications of the $Wtb$ vertex by 
non-zero values of 
the other two anomalous form factors, $f_2^+$ and $f_2^-$, 
which are often referred to as
the magnetic type anomalous couplings, are considered.
Typical values of $f_{2}^{\pm}$ are \cite{peccei}
\begin{eqnarray}
\left|f_{2}^{\pm}\right| \sim {\sqrt{m_{b}m_{t}}\over {v}}\sim 0.1.
\end{eqnarray}
They contradict neither the unitarity limit obtained from the 
$t{\bar t}$ scattering at the TeV energy scale that gives the constraint 
$\left|f_{2}^{\pm}\right|$ $\leq0.6$~\cite{renard}, nor the limits 
that are expected  from the upgraded Tevatron, which
are of order $0.2$.

The matrix elements corresponding to Eqs.~(\ref{Wtb1}) and (\ref{Wtb2}) 
have been programmed with the helicity amplitude method of \cite{KZ} 
and \cite{KJ} and then implemented into {\tt eett6f}, a Monte Carlo program 
for top quark pair production and decay into 6 fermions at linear 
colliders \cite{eett6f}. 

\section{Numerical results}

In this section, numerical results for some differential cross sections
of a selected semileptonic channel of (\ref{ee6f}) at the centre of mass 
system (CMS)
energies typical for the future linear collider with an unpolarized and 
a longitudinally polarized electron beam are shown.
The results obtained in the lowest order of SM are compared to
those obtained in the presence of the anomalous $Wtb$ coupling
of Eqs.~(\ref{Wtb1})--(\ref{rel}).
To be more specific, let us consider the angular and energy distribution 
of a $\mu^+$ and $b$-quark  of the following reaction
\bea
\label{nmdu}
 e^+e^- \rightarrow b \nu_{\mu} \mu^+\bar{b} d \bar{u}.
\eea
To the lowest order of SM, in the unitary gauge and neglecting the Higgs boson
coupling to fermion lighter than $b$-quark, reaction (\ref{nmdu}) receives 
contributions
from 264 Feynman diagrams, typical examples of which are depicted in Fig.~1. 

\begin{figure}[htb]
\centerline{
\hspace{0.5cm}
\begin{picture}(120,100)(0,0)
\ArrowLine(0,20)(30,50)
\put(5,24){\makebox(0,0)[tl]{$e^-$}}
\ArrowLine(30,50)(0,80)
\put(5,78){\makebox(0,0)[bl]{$e^+$}}
\Vertex(30,50){1}
\Photon(30,50)(60,50){2}{4}
\put(45,46){\makebox(0,0)[t]{$\gamma,Z$}}
\Vertex(60,50){1}
\ArrowLine(110,0)(85,25)
\put(100,7){\makebox(0,0)[tr]{$\bar{b}$}}
\ArrowLine(85,25)(60,50)
\put(70,28){\makebox(0,0)[br]{$\bar{t}$}}
\Photon(85,25)(100,30){-2}{3}
\Vertex(85,25){1}
\Vertex(100,30){1}
\ArrowLine(115,15)(100,30)
\put(117,17){\makebox(0,0)[l]{$\bar{u}$}}
\ArrowLine(100,30)(115,45)
\put(117,43){\makebox(0,0)[l]{$d$}}
\ArrowLine(60,50)(85,75)
\put(70,67){\makebox(0,0)[br]{$t$}}
\Vertex(85,75){1}
\Vertex(100,70){1}
\Photon(85,75)(100,70){-2}{2.5}
\put(103,67){\makebox(0,0)[tr]{\small $W^+$}}
\put(103,34){\makebox(0,0)[br]{\small $W^-$}}
\ArrowLine(115,55)(100,70)
\put(117,57){\makebox(0,0)[l]{$\mu^+$}}
\ArrowLine(100,70)(115,85)
\put(117,83){\makebox(0,0)[l]{$\nu_{\mu}$}}
\ArrowLine(85,75)(110,100)
\put(100,92){\makebox(0,0)[br]{$b$}}
\put(50,-10){\makebox(0,0)[b]{\large (a)}}
\end{picture}
\hfill
\begin{picture}(120,100)(10,0)
\ArrowLine(10,10)(45,35)
\put(20,18){\makebox(0,0)[tl]{$e^-$}}
\ArrowLine(45,35)(45,65)
\put(43,50){\makebox(0,0)[r]{$\nu_e$}}
\ArrowLine(45,65)(10,90)
\put(20,86){\makebox(0,0)[bl]{$e^+$}}
\Vertex(45,35){1}
\Vertex(45,65){1}
\Photon(45,65)(80,75){2}{4}
\put(62.5,66){\makebox(0,0)[t]{\small $W^+$}}
\Vertex(80,75){1}
\ArrowLine(110,50)(80,75)
\put(112,50){\makebox(0,0)[l]{$\bar{b}$}}
\ArrowLine(80,75)(96,85)
\put(88,82){\makebox(0,0)[br]{$t$}}
\Vertex(96,85){1}
\ArrowLine(96,85)(120,100)
\put(110,96){\makebox(0,0)[br]{$b$}}
\Photon(96,85)(110,75){-2}{2.5}
\Vertex(110,75){1}
\ArrowLine(110,75)(130,90)
\put(127,85){\makebox(0,0)[tl]{$\nu_{\mu}$}}
\ArrowLine(130,60)(110,75)
\put(127,61){\makebox(0,0)[bl]{$\mu^+$}}
\Photon(45,35)(80,25){2}{4}
\put(62.5,25){\makebox(0,0)[t]{\small $W^-$}}
\Vertex(80,25){1}
\ArrowLine(110,0)(80,25)
\put(112,0){\makebox(0,0)[l]{$\bar{u}$}}
\ArrowLine(80,25)(110,35)
\put(112,35){\makebox(0,0)[l]{$d$}}
\put(65,-10){\makebox(0,0)[b]{\large (b)}}
\end{picture}
\hfill
\begin{picture}(120,100)(0,0)
\ArrowLine(0,20)(30,50)
\put(10,28){\makebox(0,0)[tl]{$e^-$}}
\ArrowLine(30,50)(0,80)
\put(10,73){\makebox(0,0)[bl]{$e^+$}}
\Vertex(30,50){1}
\Photon(30,50)(60,50){2}{3}
\put(45,46){\makebox(0,0)[t]{$\gamma,Z$}}
\Vertex(60,50){1}
\Photon(60,50)(80,75){2}{4}
\put(75,70){\makebox(0,0)[br]{\small$W^+$}}
\Vertex(80,75){1}
\ArrowLine(110,65)(80,75)
\put(112,65){\makebox(0,0)[l]{$\mu^+$}}
\ArrowLine(80,75)(110,100)
\put(112,100){\makebox(0,0)[l]{$\nu_{\mu}$}}
\Photon(60,50)(80,25){2}{4}
\put(75,30){\makebox(0,0)[tr]{\small$W^-$}}
\Vertex(80,25){1}
\ArrowLine(80,25)(110,45)
\put(95,38){\makebox(0,0)[b]{$b$}}
\ArrowLine(120,0)(96,15)
\Vertex(96,15){1}
\ArrowLine(96,15)(80,25)
\put(86,16){\makebox(0,0)[t]{$\bar{t}$}}
\put(106,5){\makebox(0,0)[t]{$\bar{b}$}}
\Vertex(110,25){1}
\Photon(96,15)(110,25){-2}{2.5}
\ArrowLine(110,25)(130,40)
\put(125,40){\makebox(0,0)[b]{$d$}}
\ArrowLine(130,15)(110,25)
\put(125,14){\makebox(0,0)[t]{$\bar{u}$}}
\put(50,-10){\makebox(0,0)[b]{\large (c)}}
\end{picture}
\hfill
}
\vspace*{0.5cm}
\caption{Examples of the Feynman diagrams of reaction (\ref{nmdu}): 
(a) the double resonance `signal', (b) and (c) the single resonance 
diagrams.}
\end{figure}
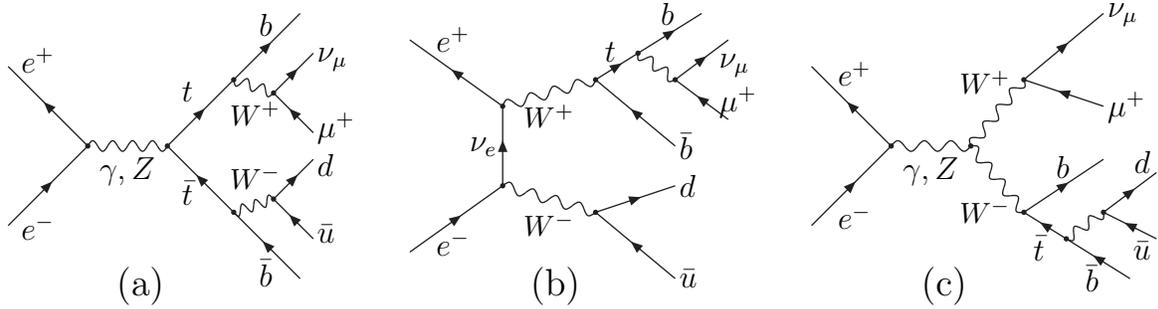

The SM electroweak physical parameters used in the computation performed
with {\tt eett6f} are the following \cite{PDG}:\\[4mm]
\centerline{
$m_W=80.419\; {\rm GeV}, \quad \Gamma_W=2.12\; {\rm GeV}, \qquad
m_Z=91.1882\; {\rm GeV}, \quad \Gamma_Z=2.4952\; {\rm GeV}$,}
\beq
\label{params}
G_{\mu}=1.16639 \times 10^{-5}\;{\rm GeV}^{-2}, \quad
m_e=0.510998902\; {\rm MeV}, \quad m_{\mu}=105.658357\; {\rm MeV},
\eeq
\centerline{
$m_u=5\; {\rm MeV}, \quad m_d=9\; {\rm MeV}, \quad  
m_t=174.3\; {\rm GeV},  \quad m_b=4.4\; {\rm GeV}$.}

The Higgs boson mass and width are assumed to be $m_H=170$ GeV and 
$\Gamma_H=0.3835$~GeV.

The SM electroweak coupling constants are given in terms of the electric 
charge $e_W=\left(4\pi\alpha_W\right)^{1/2}$ and electroweak mixing parameter 
$\sin^2\theta_W$ with
\beq
\label{alphaw}
\alpha_W=\sqrt{2} G_{\mu} m_W^2 \sin^2\theta_W/\pi, \qquad 
                \sin^2\theta_W=1-m_W^2/m_Z^2,
\eeq
where $m_W$ and $m_Z$ are physical masses of the $W^{\pm}$ and $Z^0$ boson
specified in Eq.~(\ref{params}). The strong coupling constant is given
by $g_s=\left(4\pi\alpha_s(M_Z)\right)^{1/2}$, with $\alpha_s(M_Z)=0.1181$.
The CKM mixing is neglected. Using $\sin^2\theta_W$ of Eq.~(\ref{alphaw}) 
together with the following substitutions
\beq
\label{cmass}
M_V^2=m_V^2-im_V\Gamma_V,  \quad V=W, Z,  \qquad M_H^2=m_H^2-im_H\Gamma_H,  
\qquad M_t=m_t-i\Gamma_t/2,
\eeq
which replace masses in the corresponding propagators, both in the $s$- 
and $t$-channel Feynman diagrams, is in {\tt eett6f} referred to as the
{\em `fixed width scheme'}.
The top quark width that is used in Eq.~(\ref{cmass}) is calculated to 
the lowest order
taking into account the modified $Wtb$ coupling
and performing numerical integration over the 3 particle phase space. 
Assuming $\cal{CP}$ conservation that leads to relationships
(\ref{rel}), the $Wtb$ couplings of Eq.~(\ref{Wtb1}) and Eq.~(\ref{Wtb2}) 
result in the same value, within the MC error, for the width of the top
and antitop.

Computations have been performed for different combinations of 
$f_2^+=0, \pm0.1$, $f_2^- =0, \pm 0.1$,
assuming $f_{1}^{-}=1$, $f_1^{+}=0$ and relationships (\ref{rel}). 
In the following only the results for $f_2^+=f_2^- =0.1$
are plotted, which show the biggest deviation from the SM predictions.

\begin{figure}[ht!]
\label{fig1}
\begin{center}
\setlength{\unitlength}{1mm}
\begin{picture}(35,35)(54,-50)
\rput(5.3,-6){\scalebox{0.65 0.65}{\epsfbox{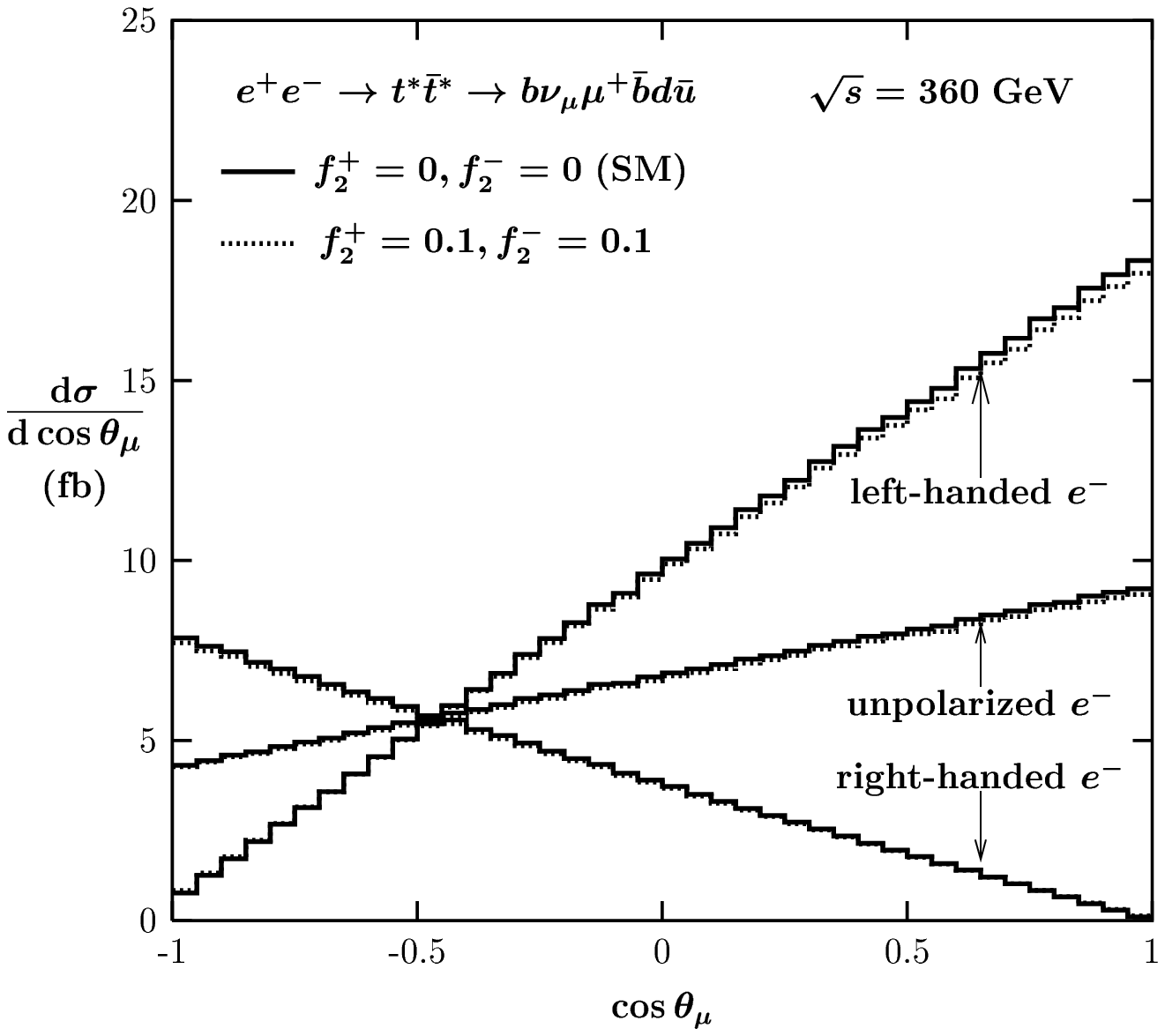}}}
\end{picture}
\begin{picture}(35,35)(9,-50)
\rput(5.3,-6){\scalebox{0.65 0.65}{\epsfbox{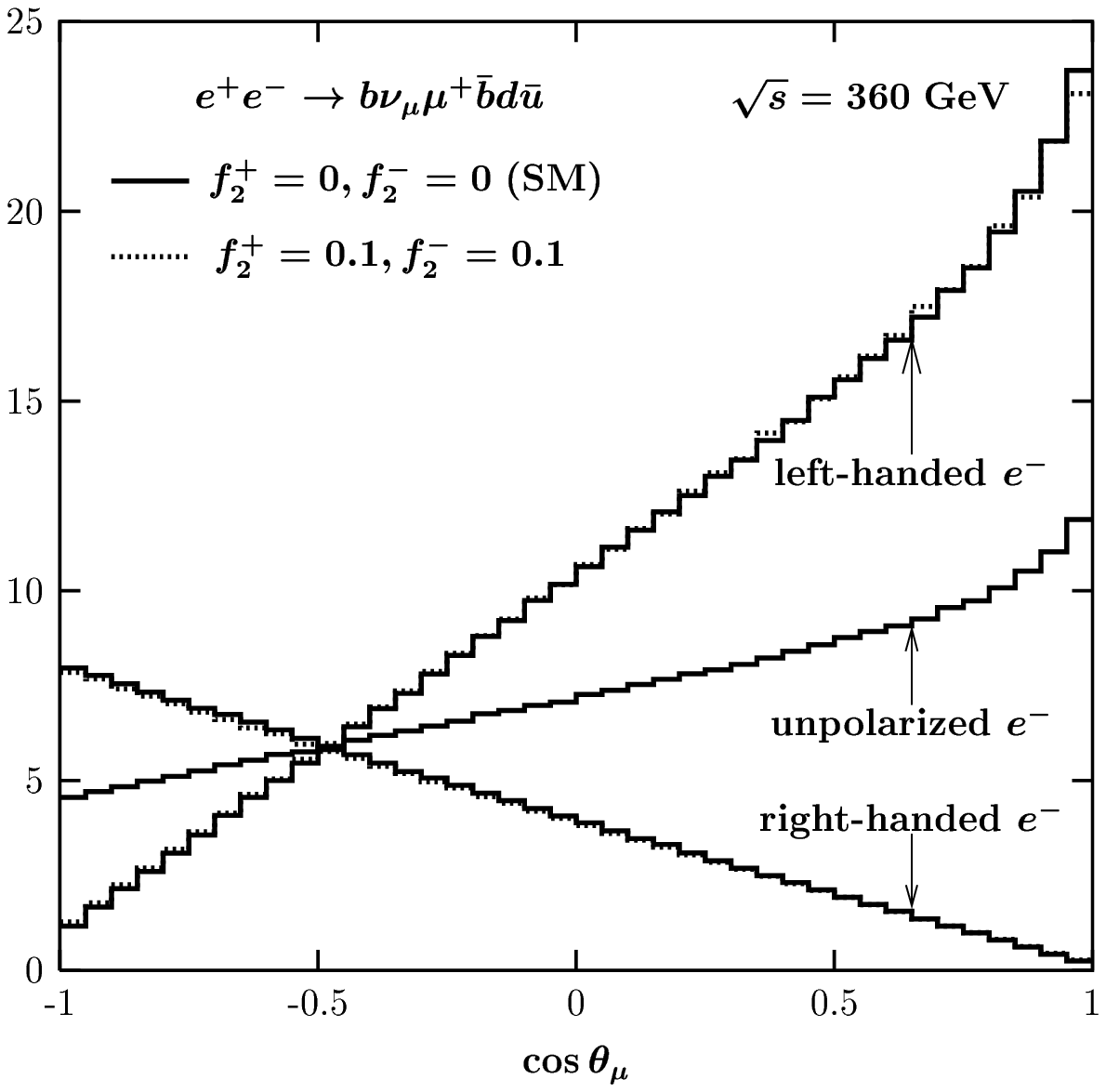}}}
\end{picture}
\end{center}
\vspace*{3.8cm}
\caption{The differential cross section ${\rm d}\sigma/{\rm d}\cos\theta_{\mu}$
         of reaction (\ref{nmdu})
         at $\sqrt{s}=360$~GeV as a function of cosine of the $\mu^+$ angle 
         with respect to the initial $e^+$ beam. The figure on the 
         left and right shows the double resonance approximation
         and the complete lowest order result, respectively.}
\end{figure}

\begin{figure}[ht!]
\label{fig2}
\begin{center}
\setlength{\unitlength}{1mm}
\begin{picture}(35,35)(54,-50)
\rput(5.3,-6){\scalebox{0.65 0.65}{\epsfbox{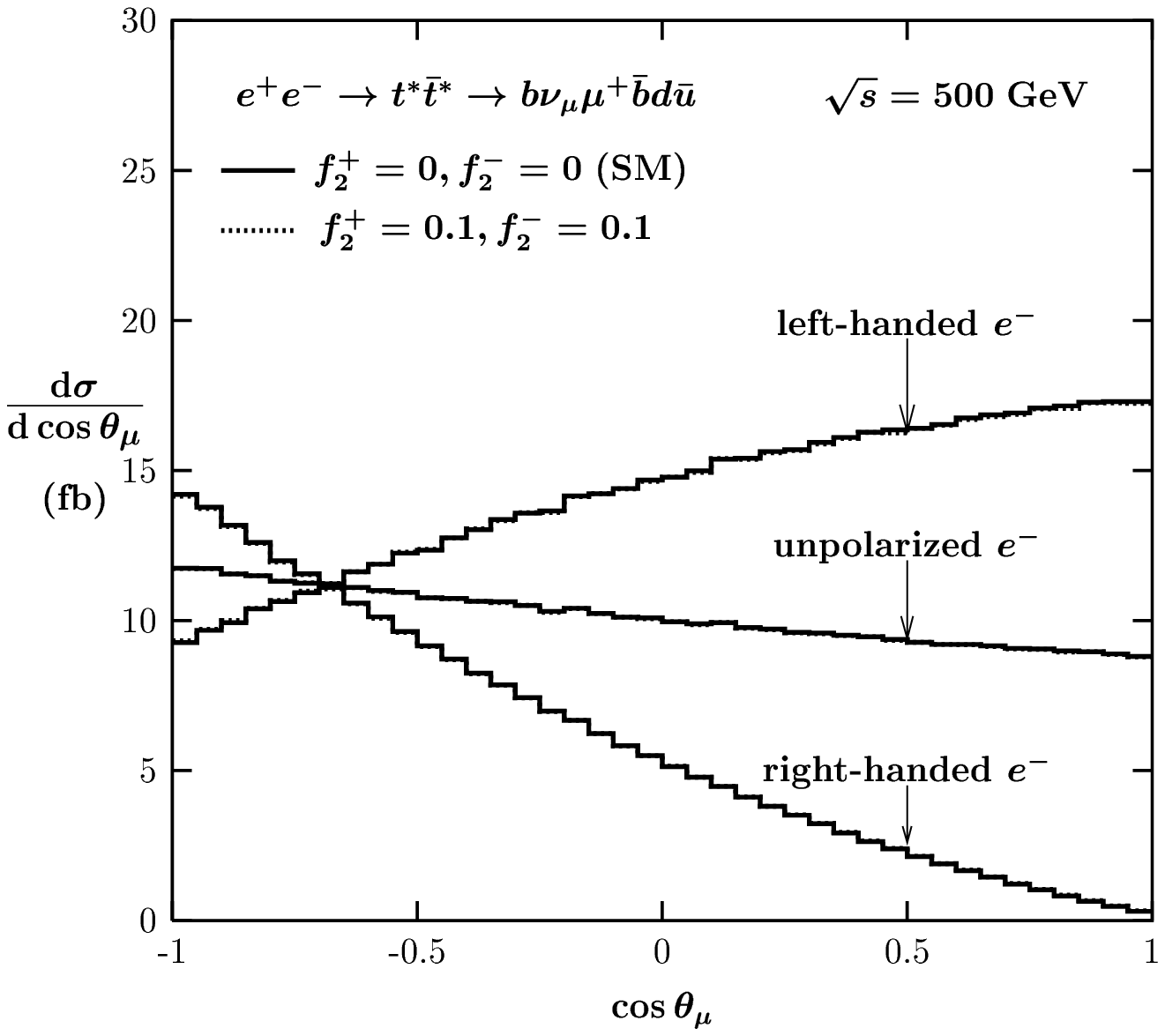}}}
\end{picture}
\begin{picture}(35,35)(9,-50)
\rput(5.3,-6){\scalebox{0.65 0.65}{\epsfbox{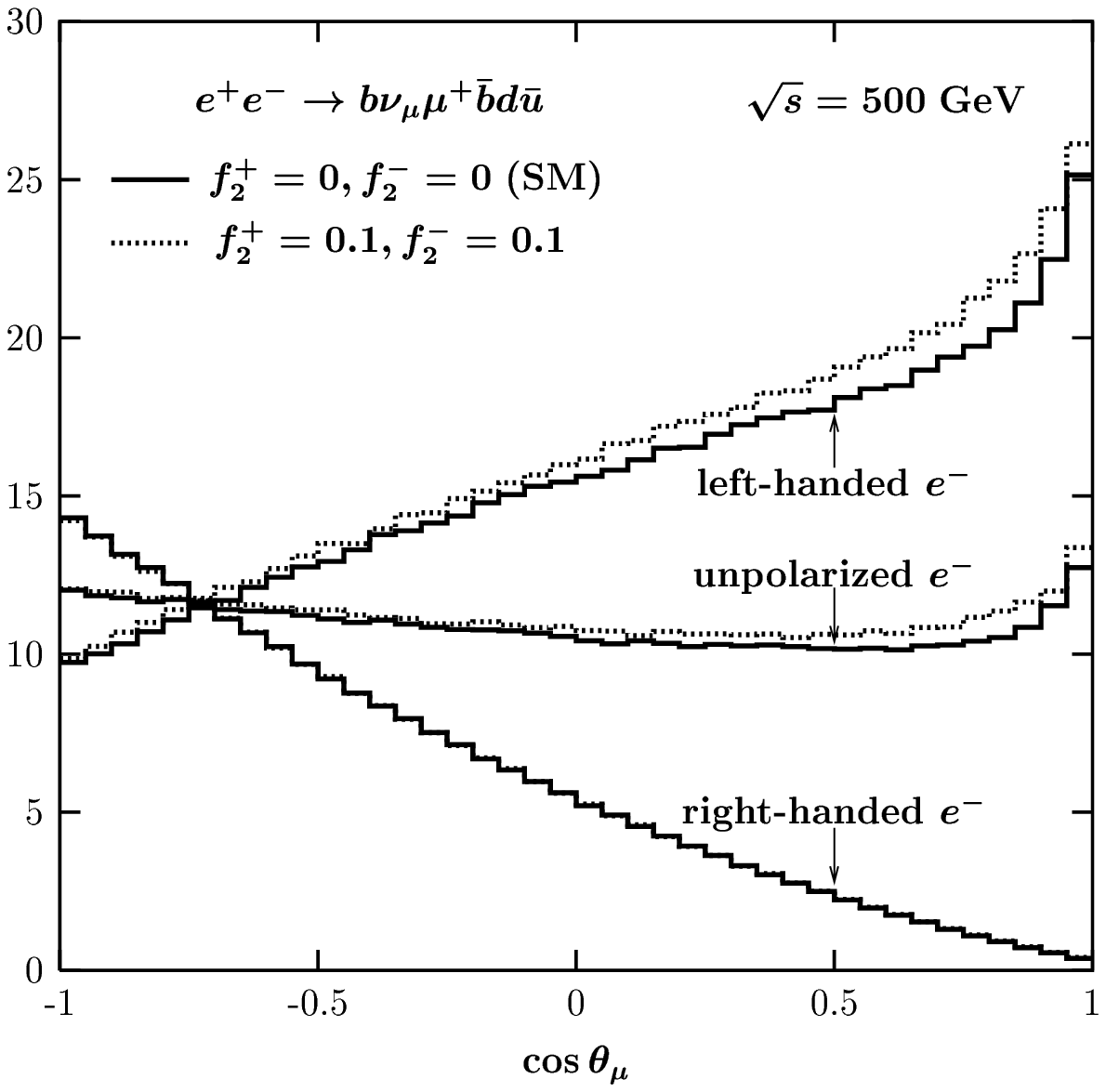}}}
\end{picture}
\end{center}
\vspace*{3.8cm}
\caption{The differential cross section ${\rm d}\sigma/{\rm d}\cos\theta_{\mu}$
         of reaction (\ref{nmdu})
         at $\sqrt{s}=500$~GeV as a function of cosine of the $\mu^+$ angle 
         with respect to the initial $e^+$ beam. The figure on the 
         left and right shows the double resonance approximation
         and the complete lowest order result, respectively.}
\end{figure}

The results for angular and energy distributions of the $\mu^+$ and
$b$-quark of reaction (\ref{nmdu}) are presented in Figs.~2--9, where 
the plots 
on the right hand side show results obtained with the complete set of 
the Feynman diagrams that contribute to (\ref{nmdu}) to the lowest order,
while the plots on the left hand side show 
results of the double resonance approximation that have been
obtained by keeping the two `signal' diagrams only, see Fig.~1a,  
and neglecting all the other Feynman diagrams of reaction (\ref{nmdu}).
In each of the figures, the solid histograms
show the SM results, while the dotted histograms show the results
in the presence of the anomalous $Wtb$ coupling.

The differential cross section ${\rm d}\sigma/{\rm d}\cos\theta_{\mu}$
at $\sqrt{s}=360$~GeV and $\sqrt{s}=500$~GeV 
is plotted in Fig.~2 and Fig.~3, respectively,
as a function of cosine of the $\mu^+$ angle with respect to the initial
$e^+$ beam.
The different histograms have been obtained with an unpolarized and 
a longitudinally polarized electron beam.
For the sake of simplicity, the level of longitudinal polarization 
is assumed to be 100\%.
Let us analyze the approximated `signal' cross sections 
on the left hand side of Fig.~2 and Fig.~3 first.
The slant of the histograms representing unpolarized cross section 
caused solely by the
Lorentz boost of the corresponding flat angular distribution
of the $\mu^+$ resulting from the decay of unpolarized top
quark at rest. The slants of the histograms representing 
polarized cross sections at $\sqrt{s}=360$~GeV, on the other hand, reflect 
proportionality of the angular distribution 
of $\mu^+$ to $(1 + \cos\theta)$, if the spin of the decaying top-quark 
points in the positive direction of the $z$ axis (spin up),
and to $(1 - \cos\theta)$, if the spin of the decaying top-quark 
points in the negative direction of the $z$ axis (spin down),
see \cite{KK1} for illustration. With the left-handedly (right-handedly) 
polarized electron beam that goes in the direction of negative $z$-axis, 
the top quark is produced preferably with its spin up (down). The corresponding
$(1 \pm \cos\theta)$ behaviour of the $\mu^+$ angular distribution
is somewhat changed by the Lorentz boost, in particular at
$\sqrt{s}=500$~GeV.
The dotted histograms represent the angular distribution
of $\mu^+$ in the presence of anomalous $Wtb$ coupling 
(\ref{Wtb1})--(\ref{rel})
with $f_1^{\pm}$ set to their SM values, $f_1^+=0$, $f_1^-=1$, and 
$f_2^+=f_2^-=0.1$. Except for a rather small effect 
in case of the left-handed electron beam at $\sqrt{s}=360$~GeV, the change 
in the $\mu^+$ angular distribution is hardly visible. This
shows that the decoupling theorem of \cite{GHR} holds in practice for reaction 
(\ref{nmdu}), even though the narrow top width approximation has not
been applied here.

The corresponding histograms on the right hand side of Fig.~2 and Fig.~3,
which have been obtained with the complete set of the Feynman diagrams 
of (\ref{nmdu}),
show some distortions, that are caused
by the non-doubly resonant background contributions. However,
they essentially show similar angular dependence as the histograms on the left.
It is interesting to note that the background enhances the anomalous effects
at $\sqrt{s}=500$~GeV, see the right hand side of Fig.~3. This means 
that due to the non-resonance background the decoupling of the anomalous
$Wtb$ coupling does not work so perfect any more.

The corresponding angular cross sections for the $b$-quark
${\rm d}\sigma/{\rm d}\cos\theta_{b}$
at $\sqrt{s}=360$~GeV and $\sqrt{s}=500$~GeV
are plotted in Fig.~4 and Fig.~5, 
respectively, as functions of cosine of the $b$-quark angle with respect 
to the initial $e^+$ beam.
Again the dotted histograms represent the cross sections
in the presence of the anomalous $Wtb$ coupling 
(\ref{Wtb1})--(\ref{rel})
with $f_1^{\pm}$ set to their SM values, $f_1^+=0$, $f_1^-=1$, and 
$f_2^+=f_2^-=0.1$. The numerical effect of the anomalous coupling
is bigger than in Figs. 2 and 3. It is visible in particular for the
longitudinally polarized electron beams.

The differential cross section ${\rm d}\sigma/{\rm d}E_{\mu}$
of reaction (\ref{nmdu}) at $\sqrt{s}=360$~GeV and $\sqrt{s}=500$~GeV
is plotted in Fig.~6 and Fig.~7, respectively, as a function of 
the $\mu^+$ energy in CMS. Again the left and right
figure show the double resonance approximation
and the complete lowest order result, respectively.
The corresponding cross sections ${\rm d}\sigma/{\rm d}E_b$
for the $b$-quark are plotted in 
Fig.~8 and Fig.~9. The anomalous effects in the energy distributions are
bigger than in the angular distribution of $\mu^+$. Note that they are
not changed by the non-resonance background.

\begin{figure}[ht!]
\label{fig3}
\begin{center}
\setlength{\unitlength}{1mm}
\begin{picture}(35,35)(54,-50)
\rput(5.3,-6){\scalebox{0.65 0.65}{\epsfbox{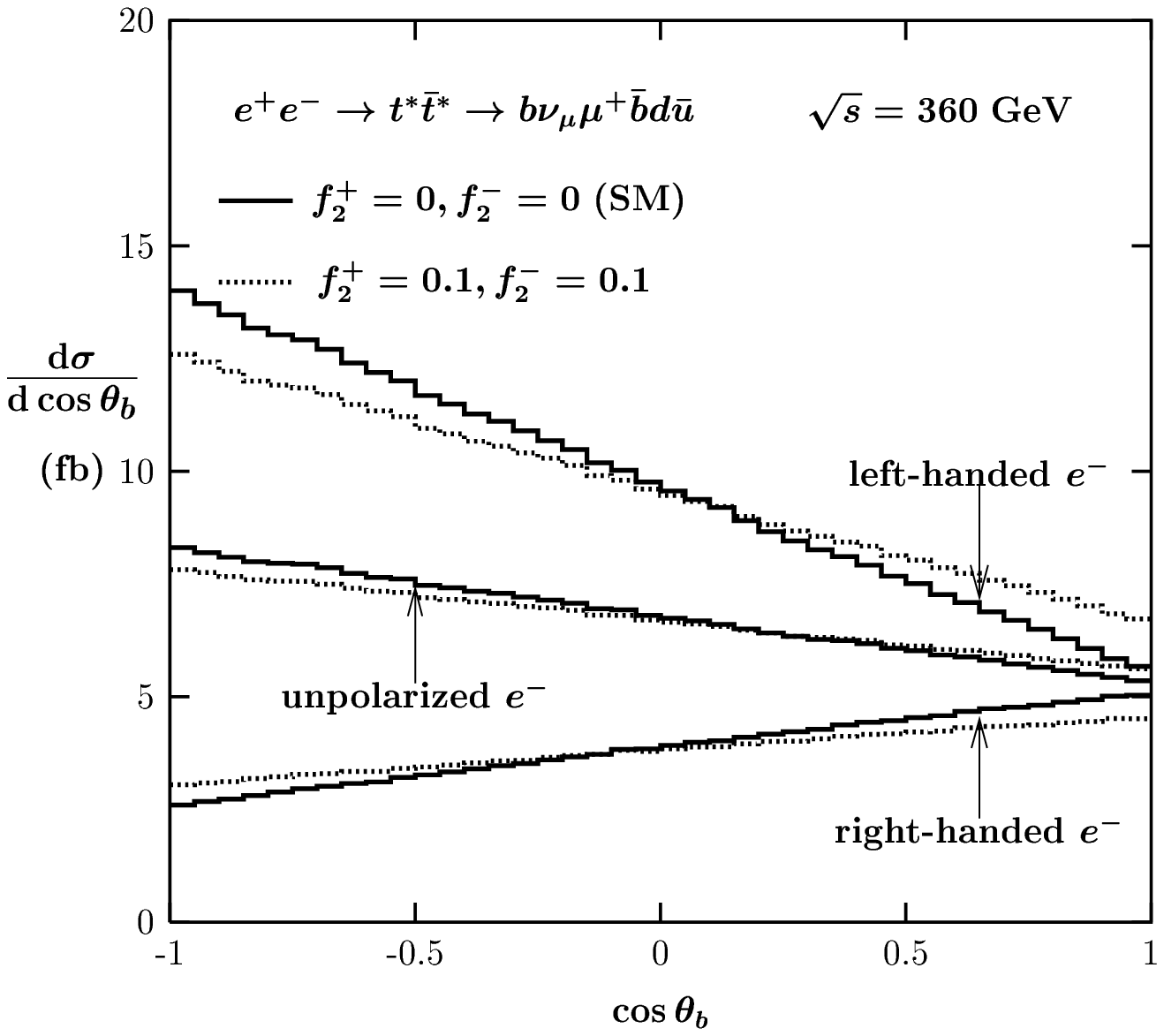}}}
\end{picture}
\begin{picture}(35,35)(9,-50)
\rput(5.3,-6){\scalebox{0.65 0.65}{\epsfbox{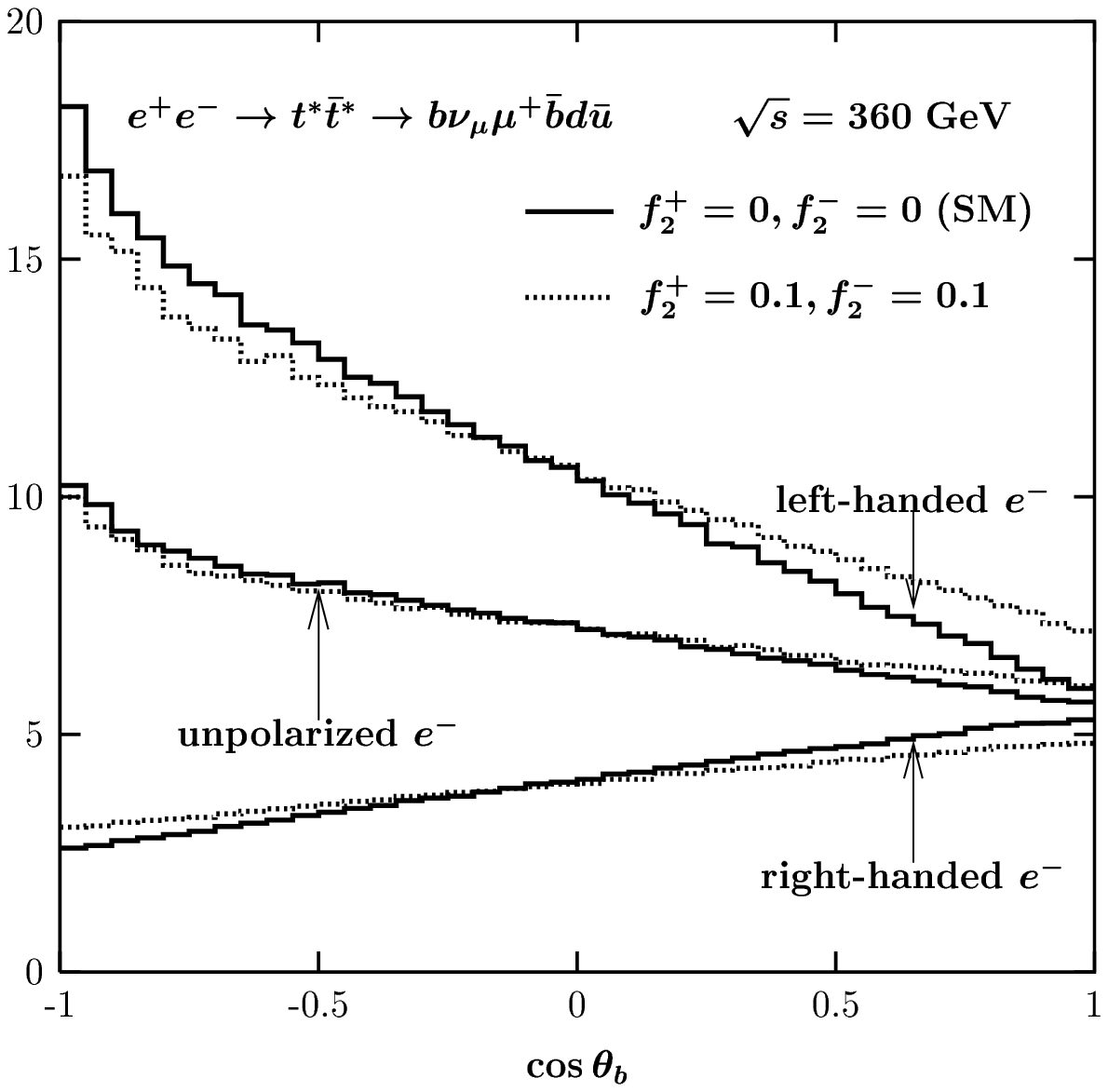}}}
\end{picture}
\end{center}
\vspace*{3.8cm}
\caption{The differential cross section ${\rm d}\sigma/{\rm d}\cos\theta_b$
         of reaction (\ref{nmdu})
         at $\sqrt{s}=360$~GeV as a function of cosine of the $b$-quark angle 
         with respect to the initial $e^+$ beam. The figure on the 
         left and right shows the double resonance approximation
         and the complete lowest order result, respectively.}
\end{figure}

\begin{figure}[ht!]
\label{fig4}
\begin{center}
\setlength{\unitlength}{1mm}
\begin{picture}(35,35)(54,-50)
\rput(5.3,-6){\scalebox{0.65 0.65}{\epsfbox{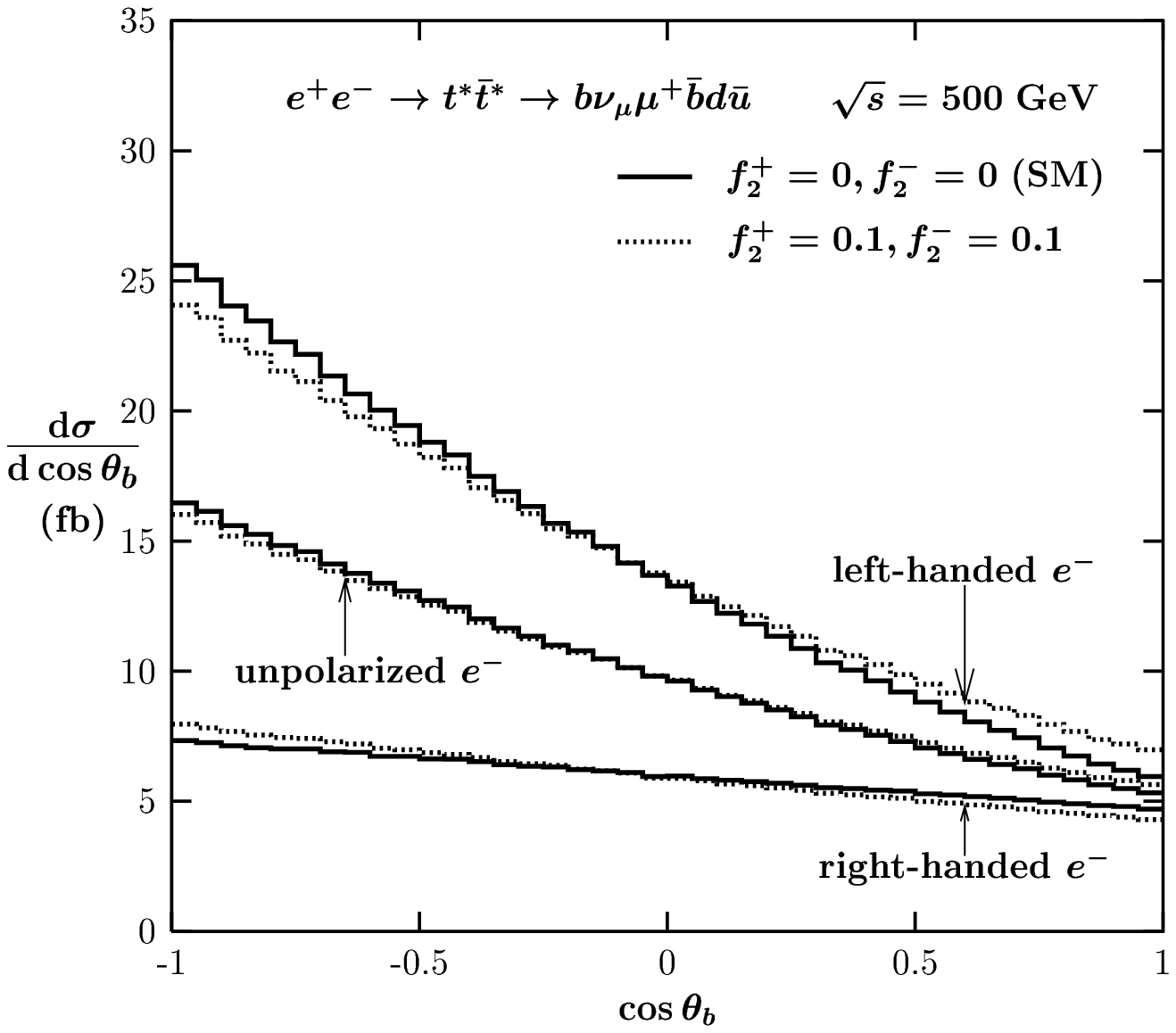}}}
\end{picture}
\begin{picture}(35,35)(9,-50)
\rput(5.3,-6){\scalebox{0.65 0.65}{\epsfbox{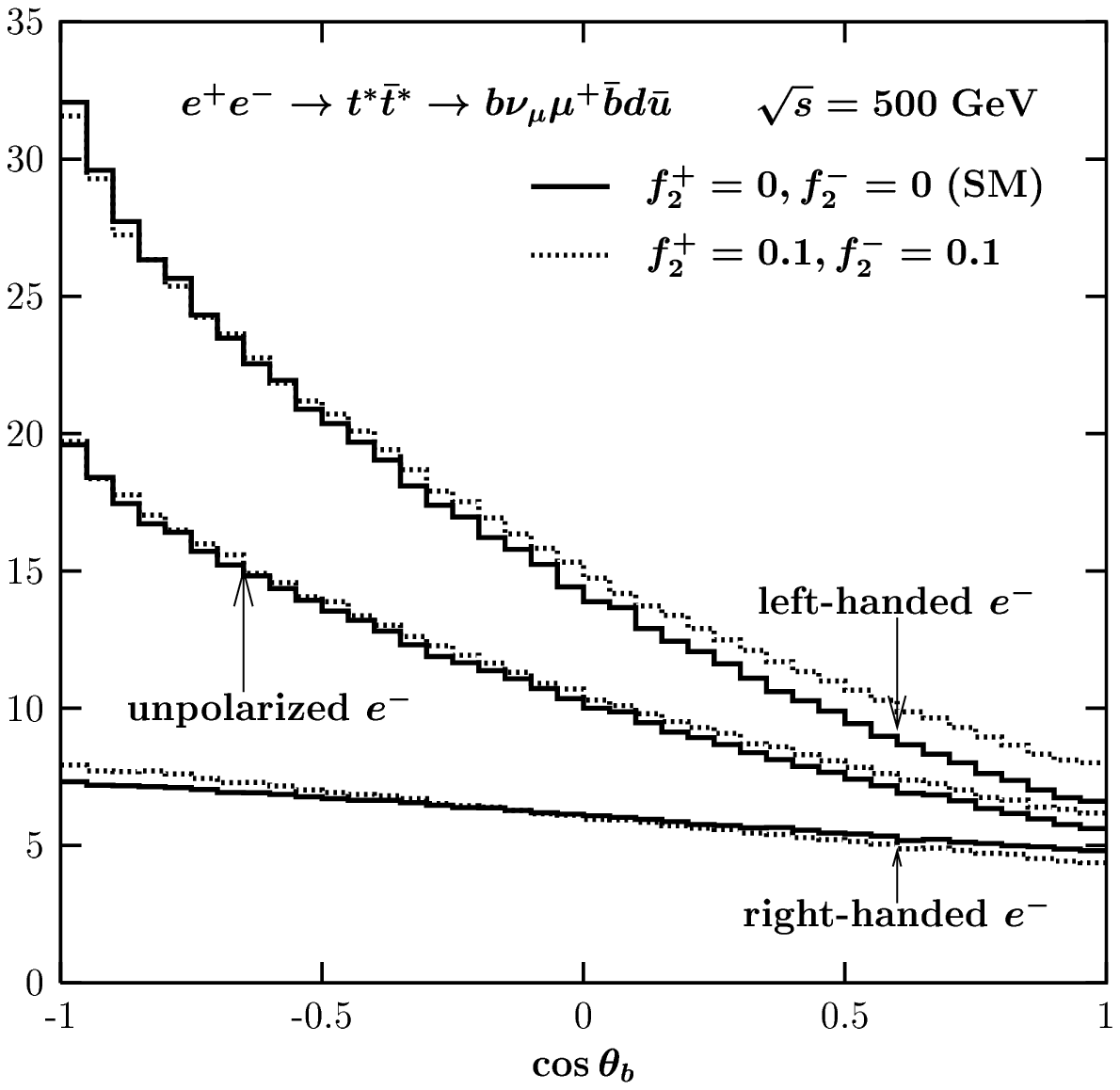}}}
\end{picture}
\end{center}
\vspace*{3.8cm}
\caption{The differential cross section ${\rm d}\sigma/{\rm d}\cos\theta_b$
         of reaction (\ref{nmdu})
         at $\sqrt{s}=500$~GeV as a function of cosine of the $b$-quark angle 
         with respect to the initial $e^+$ beam. The figure on the 
         left and right shows the double resonance approximation
         and the complete lowest order result, respectively.}
\end{figure}

\begin{figure}[ht!]
\label{fig5}
\begin{center}
\setlength{\unitlength}{1mm}
\begin{picture}(35,35)(54,-50)
\rput(5.3,-6){\scalebox{0.65 0.65}{\epsfbox{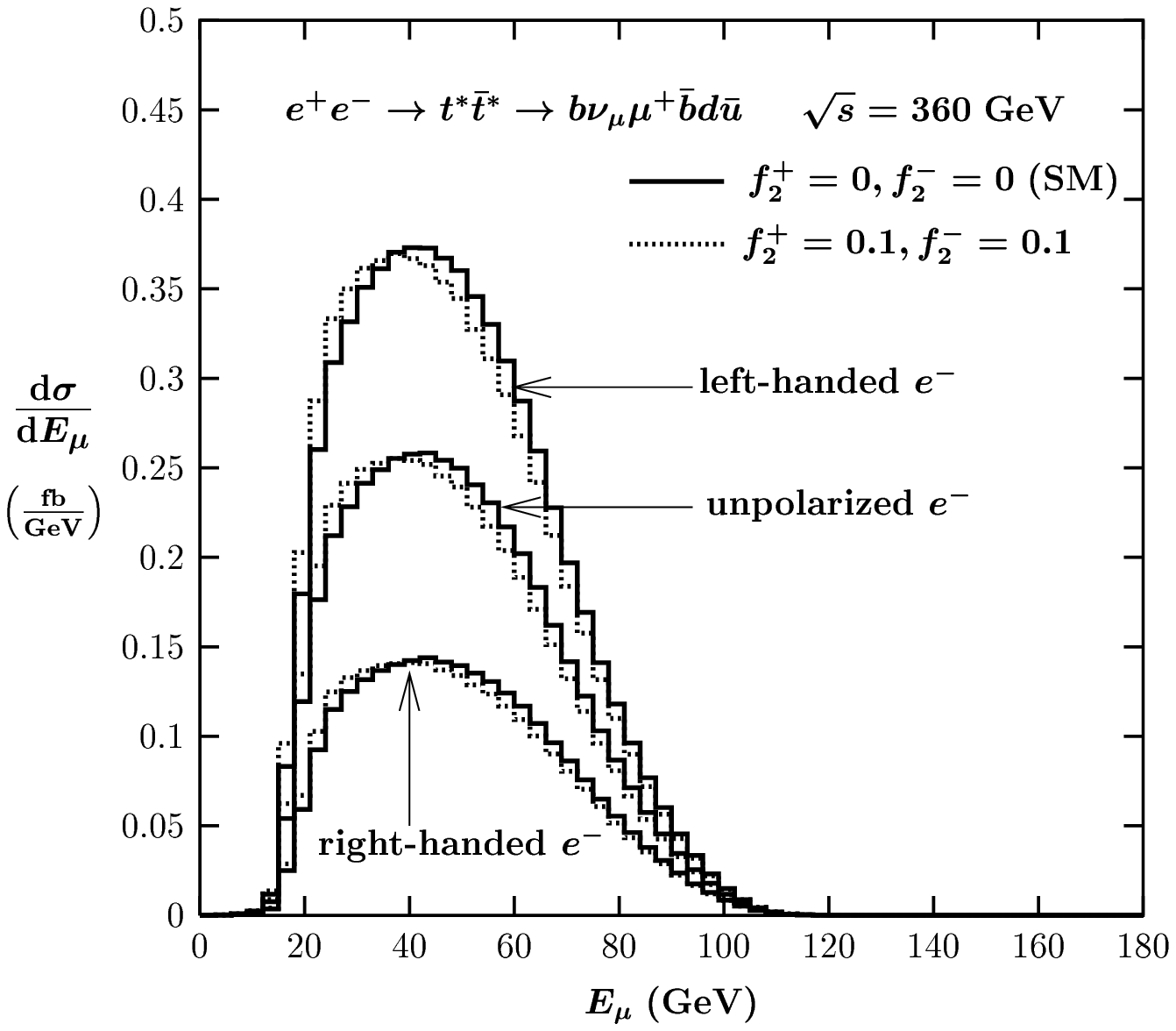}}}
\end{picture}
\begin{picture}(35,35)(9,-50)
\rput(5.3,-6){\scalebox{0.65 0.65}{\epsfbox{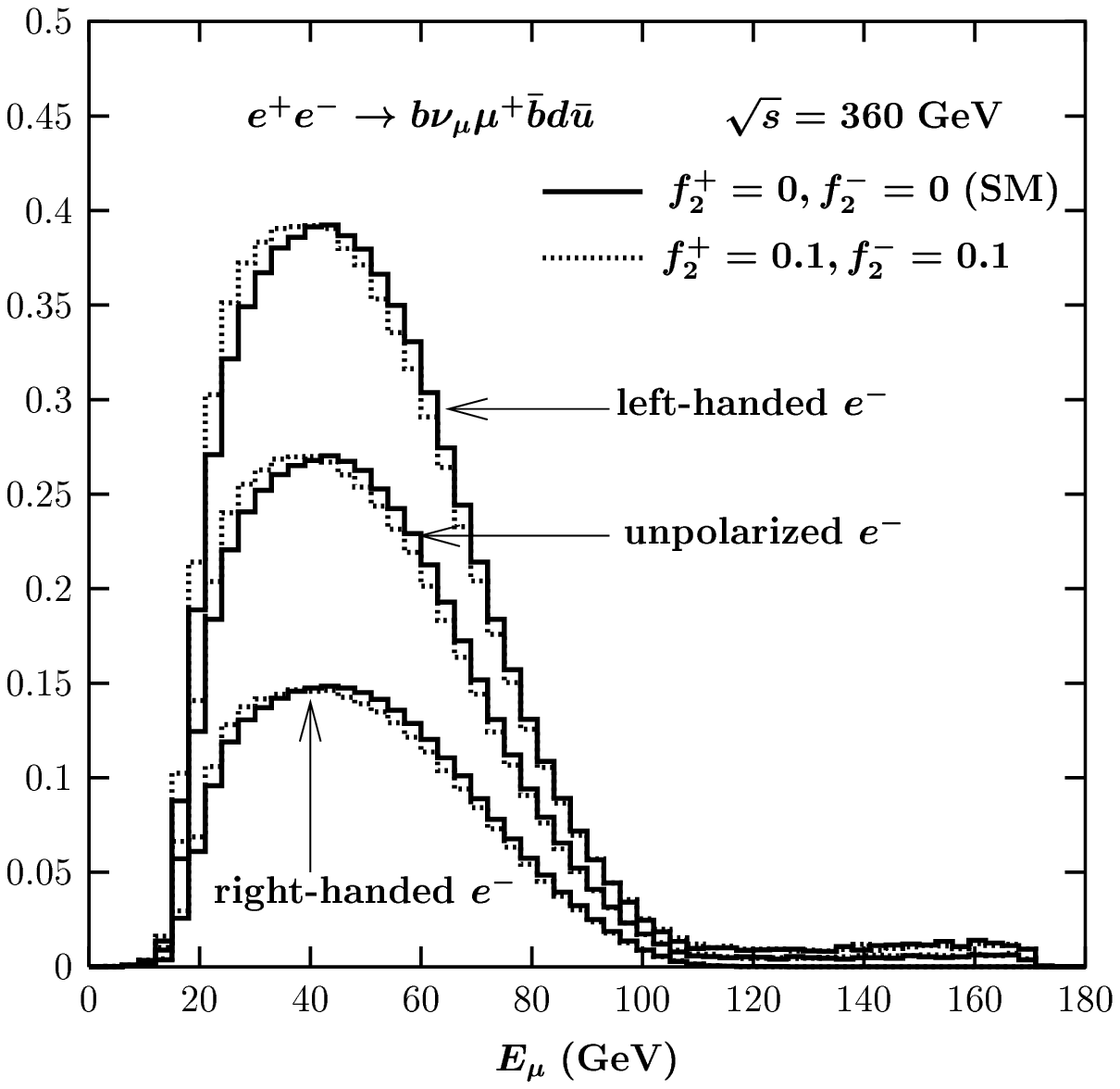}}}
\end{picture}
\end{center}
\vspace*{3.8cm}
\caption{The differential cross section ${\rm d}\sigma/{\rm d}E_{\mu}$
         of reaction (\ref{nmdu})
         at $\sqrt{s}=360$~GeV as a function of the $\mu^+$ energy
         in CMS. The figure on the 
         left and right shows the double resonance approximation
         and the complete lowest order result, respectively.}
\end{figure}

\begin{figure}[ht!]
\label{fig6}
\begin{center}
\setlength{\unitlength}{1mm}
\begin{picture}(35,35)(54,-50)
\rput(5.3,-6){\scalebox{0.65 0.65}{\epsfbox{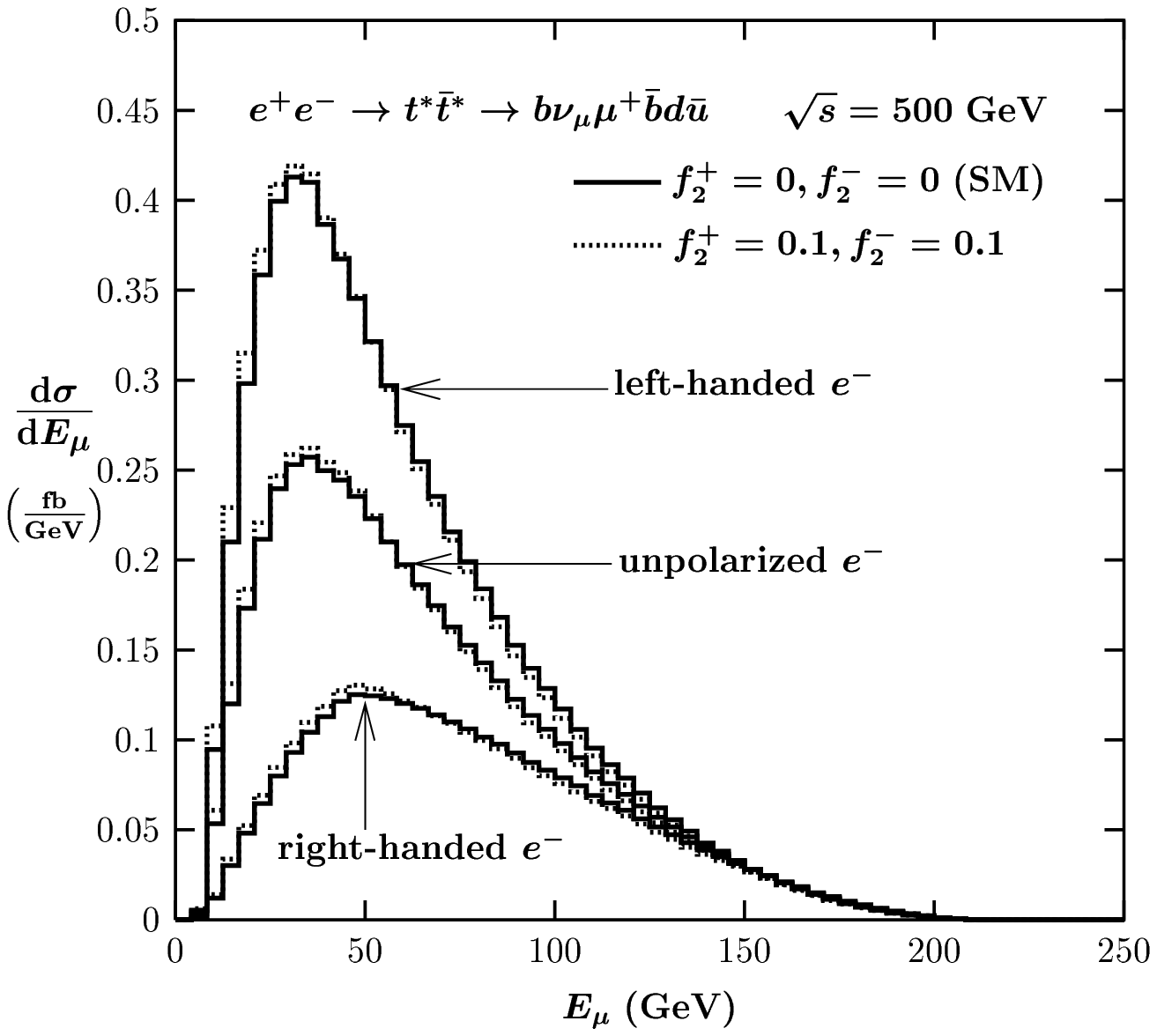}}}
\end{picture}
\begin{picture}(35,35)(9,-50)
\rput(5.3,-6){\scalebox{0.65 0.65}{\epsfbox{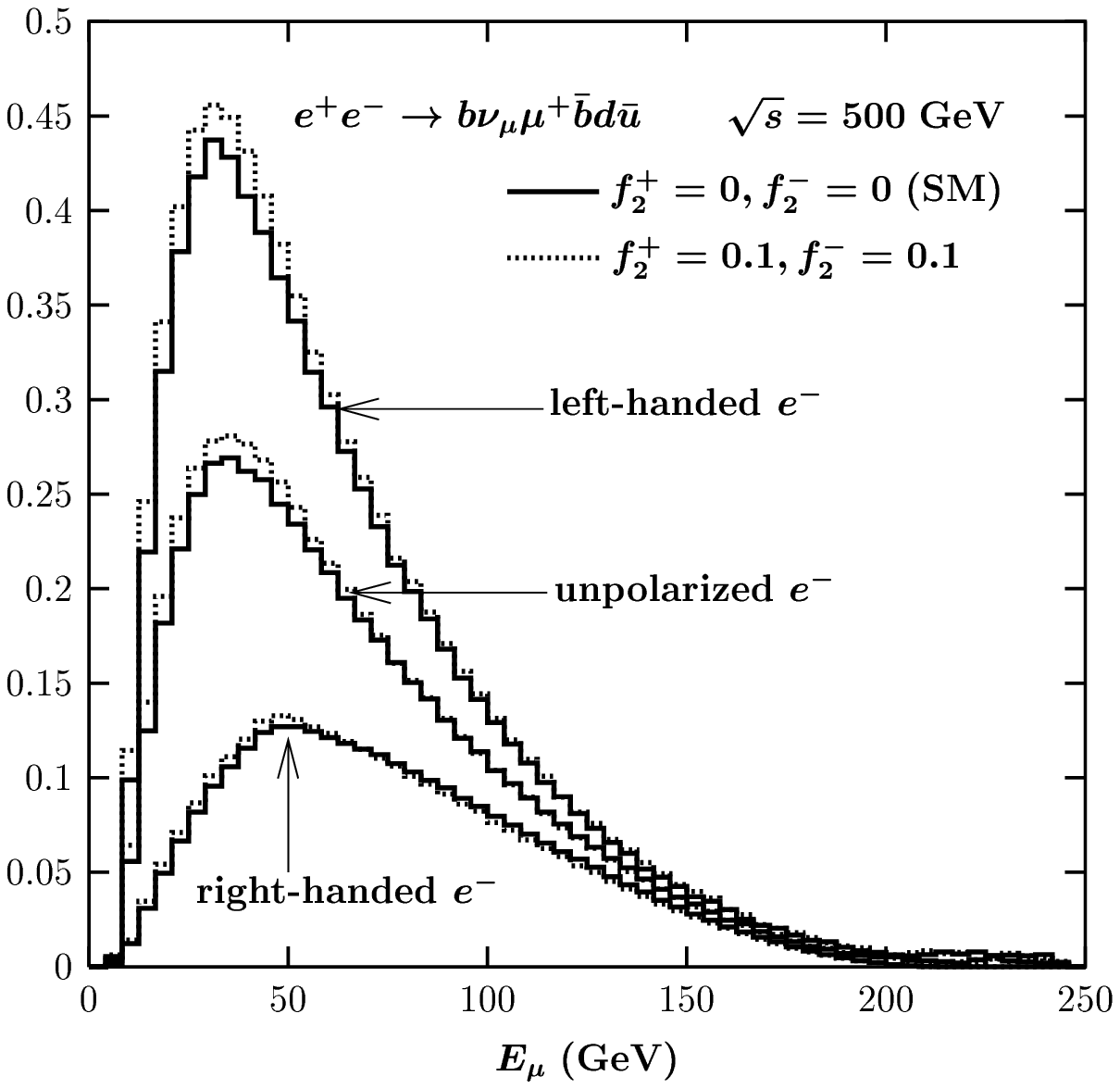}}}
\end{picture}
\end{center}
\vspace*{3.8cm}
\caption{The differential cross section ${\rm d}\sigma/{\rm d}E_{\mu}$
         of reaction (\ref{nmdu})
         at $\sqrt{s}=500$~GeV as a function of the $\mu^+$ energy
         in CMS. The figure on the 
         left and right shows the double resonance approximation
         and the complete lowest order result, respectively.}
\end{figure}

\begin{figure}[ht!]
\label{fig7}
\begin{center}
\setlength{\unitlength}{1mm}
\begin{picture}(35,35)(54,-50)
\rput(5.3,-6){\scalebox{0.65 0.65}{\epsfbox{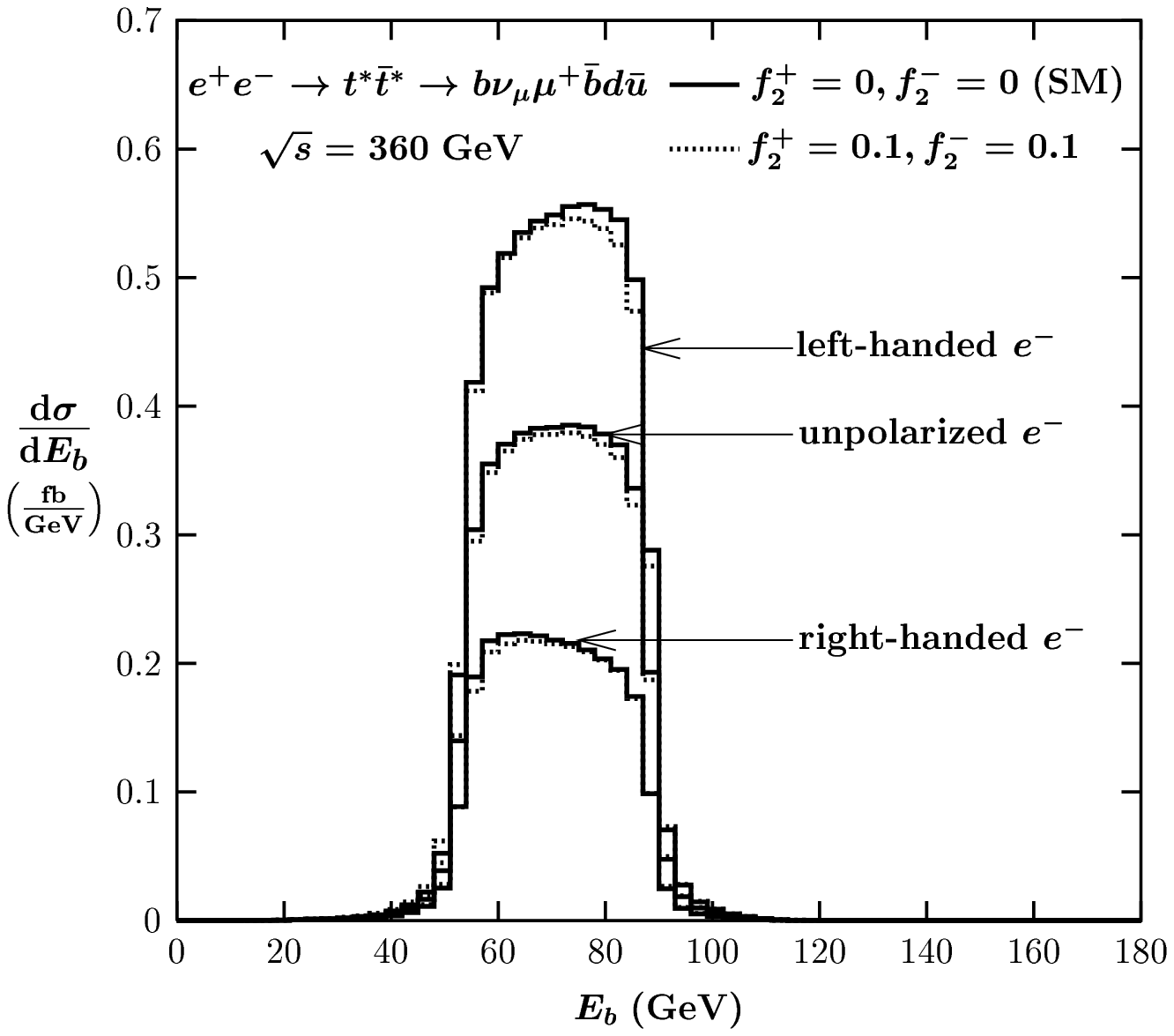}}}
\end{picture}
\begin{picture}(35,35)(9,-50)
\rput(5.3,-6){\scalebox{0.65 0.65}{\epsfbox{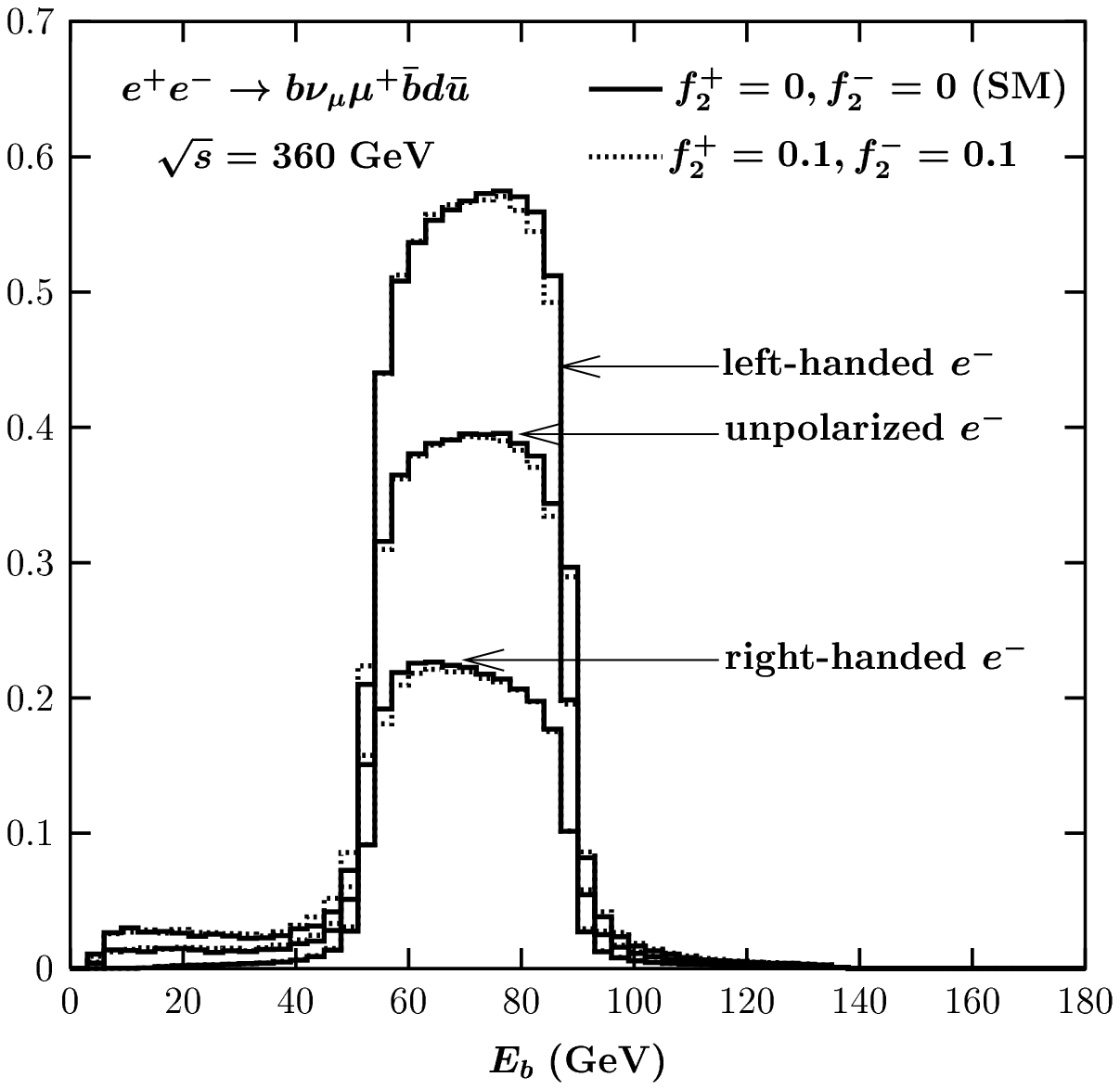}}}
\end{picture}
\end{center}
\vspace*{3.8cm}
\caption{The differential cross section ${\rm d}\sigma/{\rm d}E_b$
         of reaction (\ref{nmdu})
         at $\sqrt{s}=360$~GeV as a function of the $b$-quark energy
         in CMS. The figure on the 
         left and right shows the double resonance approximation
         and the complete lowest order result, respectively.}
\end{figure}

\begin{figure}[ht!]
\label{fig8}
\begin{center}
\setlength{\unitlength}{1mm}
\begin{picture}(35,35)(54,-50)
\rput(5.3,-6){\scalebox{0.65 0.65}{\epsfbox{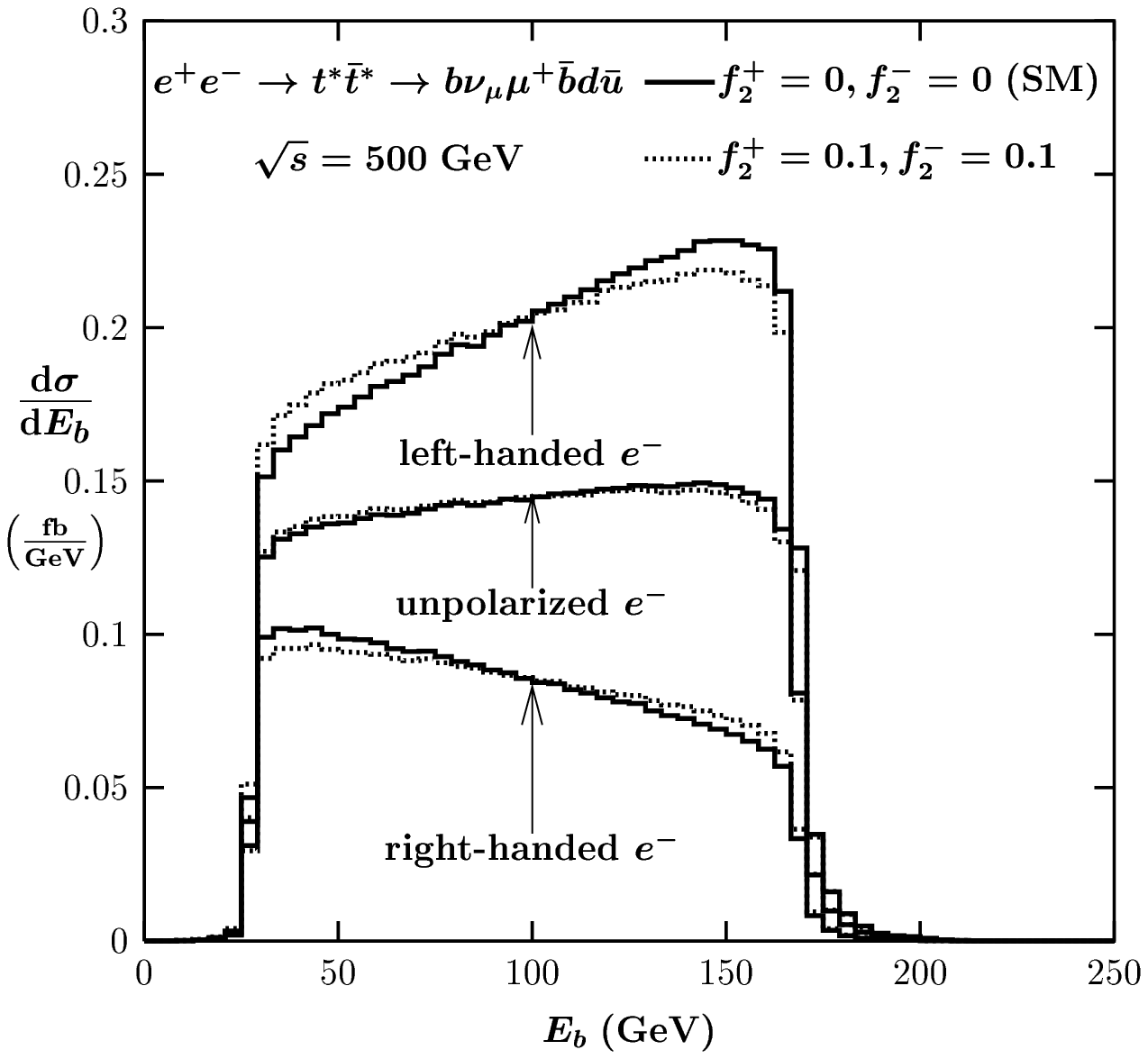}}}
\end{picture}
\begin{picture}(35,35)(9,-50)
\rput(5.3,-6){\scalebox{0.65 0.65}{\epsfbox{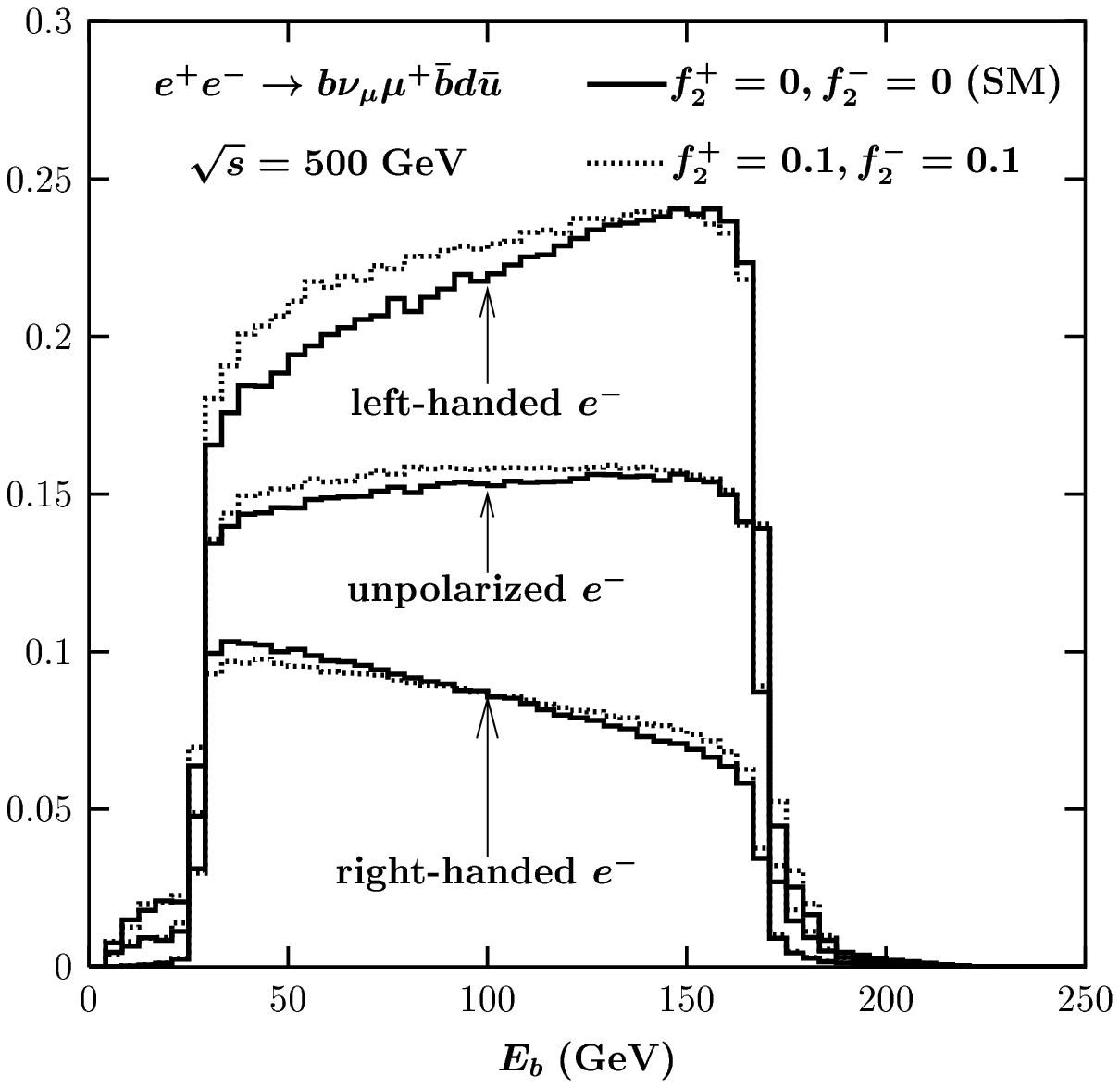}}}
\end{picture}
\end{center}
\vspace*{3.8cm}
\caption{The differential cross section ${\rm d}\sigma/{\rm d}E_b$
         of reaction (\ref{nmdu})
         at $\sqrt{s}=500$~GeV as a function of the $b$-quark energy
         in CMS. The figure on the 
         left and right shows the double resonance approximation
         and the complete lowest order result, respectively.}
\end{figure}

\section{Summary}

Results on angular and energy distributions of a $\mu^+$ and 
$b$-quark of reaction (\ref{nmdu}), which is a typical semileptonic
channel of the top quark pair production 
at a future linear collider,
for the collisions of
the unpolarized and longitudinally polarized electron beam
against the unpolarized positron beam at $\sqrt{s}=360$~GeV and 
$\sqrt{s}=500$~GeV, have been presented.
The results, which have been computed to the lowest order
in the SM and in the model with the anomalous $Wtb$ coupling,
taking into account the complete set of the lowest order Feynman diagrams 
of (\ref{nmdu}) and the two top quark pair production signal diagrams 
of Fig.~1a alone, 
illustrate how the anomalous $Wtb$ coupling modifies the SM results.
In particular, Figs. 2 and 3 illustrate that the angular distribution of 
$\mu^+$ receives practically no contribution from the anomalous 
$Wtb$ coupling. This shows that the decoupling theorem that has been 
proven in literature \cite{GHR} in the narrow top quark width approximation 
holds in practice also in a more realistic
case, where the top quark pair is produced off shell and 
the non-doubly resonant background contributions are taken into account. 
Analysis of the $\mu^+$ distributions obtained with the longitudinally 
polarized beam shows that they are a very sensitive probe of the top quark
polarization, as expected.

\end{document}